\documentclass[aps,pra,reprint,nofootinbib,superscriptaddress,twocolumn,showpacs,showkeys,longbibliography,amsmath,amssymb]{revtex4-1}
\usepackage{graphicx}
\usepackage{dcolumn}
\usepackage{bm}
\usepackage{braket}
\usepackage{subfigure}
\usepackage[colorlinks,bookmarks=false,citecolor=blue,linkcolor=red,urlcolor=blue]{hyperref}
\usepackage[english]{babel}
\usepackage{changes}
\usepackage{CJK}

\begin{document}
\preprint{APS/123-QED}

\title{Hierarchical dimensional crossover of an optically-trapped quantum gas with disorder}
\author{Kangkang Li }
\affiliation{Department of Physics, Zhejiang Normal University, Jinhua 321004, China}
\author{Zhaoxin Liang }\email[Corresponding author:~] {zhxliang@zjnu.edu.cn}
\affiliation{Department of Physics, Zhejiang Normal University, Jinhua 321004, China}

\date{\today}
\begin{abstract}
Dimensionality serves as an indispensable ingredient in any attempt to formulate the low-dimensional physics, and studying the dimensional crossover
at a fundamental level is challenging. The purpose of this work is to study the hierarchical  dimensional crossovers, namely the crossover from three dimensions (3D) to quasi-2D and then to 1D. Our system consists of a 3D Bose-Einstein condensate (BEC) trapped in an anisotropic 2D optical lattice characterized by the lattice depths $V_1$ along the $x$ direction and $V_2$ along the $y$ direction, respectively, where the hierarchical dimensional crossover is controlled via $V_1$ and $V_2$. We analytically derive the ground-state energy, quantum depletion and the superfluid density of the system. Our results demonstrate the 3D-quasi-2D-1D dimensional crossovers in the behavior of quantum fluctuations. Conditions for possible experimental realization of our scenario are also discussed.
\end{abstract}

\maketitle


{\it Introduction.---}
Dimensionality plays a fundamental role in determining the properties of quantum many-body systems. It underpins many remarkable phenomena such as the high-Tc superconductivity~\cite{Lee2006} and magic-angle graphene~\cite{Cao2018a,Cao2018b,Tarnopolsky2019} in two dimensions (2D) and the Tomonaga-Luttinger liquid~\cite{Haldane1981} in 1D. Therefore, there are ongoing interests and great efforts in investigating how dimensionality affects the properties of quantum many-body systems.

Tightly confined Bose-Einstein condensate (BEC)~\cite{Bloch2008} provides an ideal playground for the theoretical and experimental explorations of the dimensional effects. In particular, the state-of-the-art technology allows the depth of an optical lattice to be arbitrarily tuned by changing the laser intensities, enabling realizations of quasi-1D~\cite{Paredes2004} and quasi-2D~\cite{Peppler2018,Holten2018} BECs. Thus, an important direction of investigation consists of studying the properties of a BEC system in the dimensional crossover.

Along this research line, many researches have been carried out. For instance, Refs.~\cite{Orso2006,Hu2009} have shown that the presence of a 2D lattice
can induce a 3D to 1D crossover in the behavior
of quantum fluctuations; Refs.~\cite{Hu2011,Zhou2010,Faigle2021} have investigated quantum phases along the 3D-2D crossover and the visualization of the dimensional effects in collective excitations.
These works~\cite{Orso2006,Hu2009,Hu2011,Zhou2010,Hu2019,Yin2020,Faigle2021,Yao2022} consider the tight confinement scheme that gives rise to a direct dimensional crossover from 3D to 2D or 1D (i.e., 3D-2D or 3D-1D crossover). Instead, we will be interested in the \textit{hierarchical} dimensional crossovers, i.e., the 3D-quasi-2D-1D dimensional crossovers.

We will be interested in the effect of dimensionality on not only the ground state energy and quantum depletion but also the transport properties. This is motivated by experimental realizations of BECs in the presence of disorder~\cite{White2009,Paiva2015}. For example, superfluidity represents a kinetic property of a system, and the superfluid density is a transport coefficient determined by
the linear response theory. In this work, we investigate the 3D - quasi-2D - 1D crossovers in the properties of a disorder BEC trapped in an anisotropic optical lattice, using the Green function approach. Specifically we calculate the ground-state energy and quantum depletion, as well as the superfluid density,  and we analyze the combined effects of dimensionality and disorder.

{\it Model.---}  At zero temperature, an optically-trapped BEC can be well described by the $N$-body Hamiltonian~\cite{Orso2006,Hu2009,Hu2011,Zhou2010}
\begin{eqnarray}
   \hat{H}-\mu \hat{N}=\int {\rm d}\bm{r}\hat{\Psi}^\dagger(\bm{r})\bigg[-\frac{\hbar^2\nabla^2}{2m}-\mu
   &+&V_{\text{opt}}(\bm{r})+V_{\text{ran}}(\bm{r})\nonumber\\
   &+&\frac{g}{2}\hat{\Psi}^\dagger\hat{\Psi}\bigg]
   \hat{\Psi}(\bm{r}),\label{Model}
\end{eqnarray}
where $\hat{\Psi}(\bm{r})$ is the field operator for bosons with mass $m$, $\mu$ is the chemical potential, $\hat{N}=\int {\rm d}\bm {r}\hat{\Psi}^\dagger(\bm {r})\Psi(\bm {r})$
is the number operator, and $g=4\pi\hbar^2 a_{\text 3D }/m$ is the coupling constant with $a_{\text 3D}$ being the 3D scattering length~\cite{Petrov2000}. In Hamiltonian~(\ref{Model}), $V_{\text{opt}}(\bm{r})$ and  $V_{\text{ran}}(\bm{r})$, respectively, describe the anisotropic 2D optical lattice and the external random potential.

We consider the anisotropic 2D optical lattice in Hamiltonian (\ref{Model}) in the form~\cite{Bloch2008}
\begin{equation}
V_{\text{opt}}(\bm{r})=E_R[V_1\sin^2(q_Bx)+V_2\sin^2(q_By)],\label{OL}
\end{equation}
where $V_{1,2}$ denote the laser intensities and $E_R=\hbar^2q_B^2/2m$ is the recoil energy, with $\hbar q_B$ being the Bragg momentum and $m$ the atomic mass. The lattice period is fixed by $d=\pi/q_B$. Atoms are free in the $z$ direction. By controlling the depths of the optical lattice $V_1$ and $V_2$, crossovers to low dimensions are expected to occur via the hierarchical access of new energy scales: firstly, a 3D Bose gas
becomes quasi-2D when the energetic restriction to freeze $x$-direction excitations is reached; next, by further freezing the kinetic energy along the $y$-direction, the quasi-2D BEC is expected to enter the quasi-1D regime.

Moreover, in Hamiltonian (\ref{Model}), $V_{\text{ran}}(\bm{r})=\sum_i^{N_{\text{imp}}}v(\bm {r}-\bm {r}_i)$ can be produced by the random potential~\cite{Huang1992,Hu2009,Zhou2010}. For sufficiently dilute disorder~\cite{White2009,Paiva2015},
$v(\bm {r})$ can be approximated by an effective pseudopotential, i.e., $v(\bm {r})=g_{\text{imp}}\delta (\bm {r})$, where $g_{\rm{imp}}=2\pi \hbar^2 \tilde{b}/m$ is the effective
coupling constant of an impurity-boson pair and $\tilde{b}$ is the effective scattering length accounting for the presence of a 2D optical lattice~\cite{Astrakharchik2002,Yao2020}.

\begin{figure}
      \includegraphics[width=0.5\textwidth]{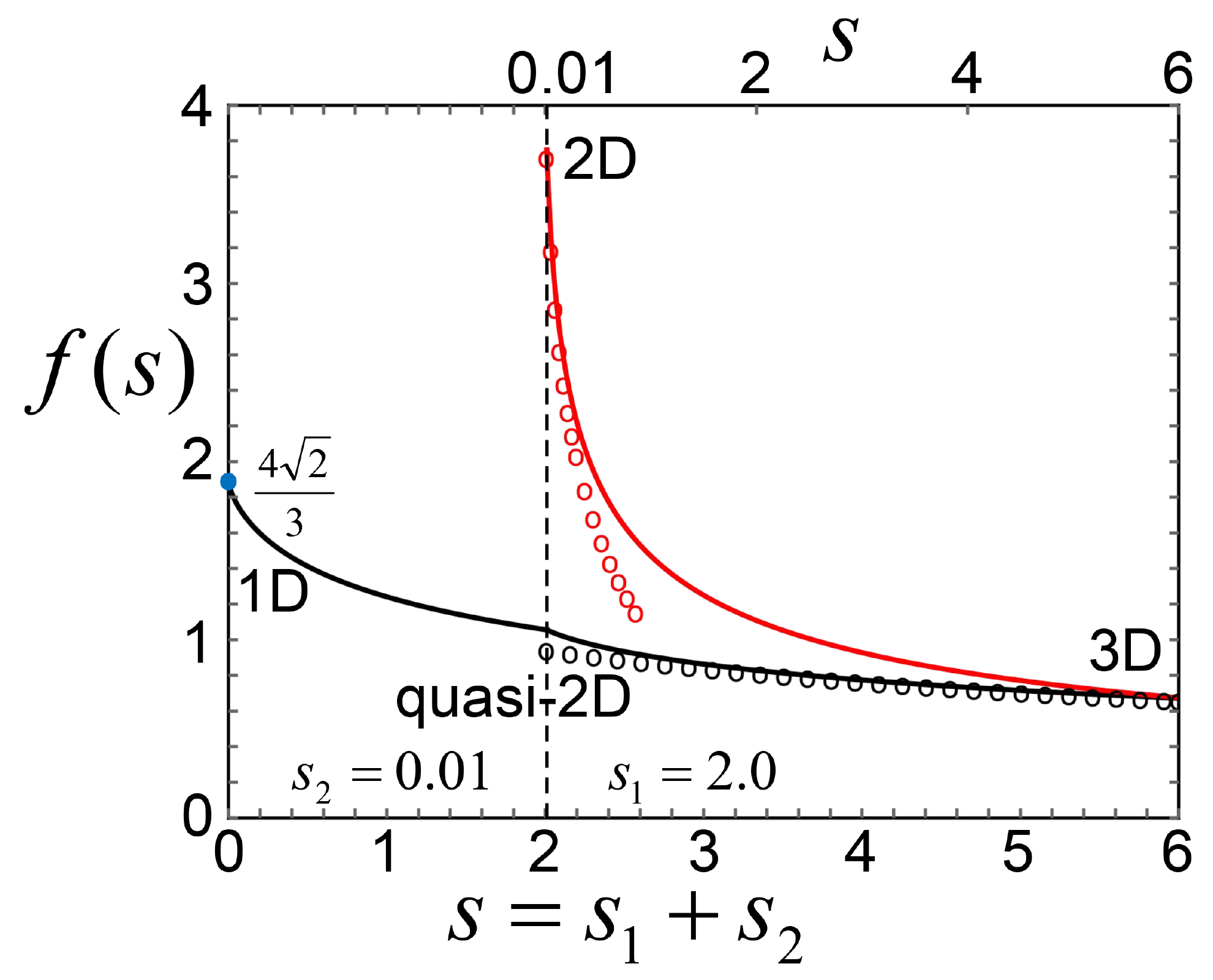}
      \caption{Scaling function $f(s)$ in Eq. (\ref{ffun}) (black solid line) and its 3D (black dotted line) asymptotic behavior with $s=s_1+s_2$ being the dimensionless tunneling rates. In the left side of the vertical dashed line, we fix $s_2=0.01$ and set $s_1=[0,2]$. The blue dot denotes the 1D Lieb-Liniger limit of $f(0)=\frac{4\sqrt{2}}{3}$. In the right side, we fix $s_1=2$ and set $s_2=[0.01,4]$. As $s$ decreases from $6$ to $0$, the model system realizes the step-by-step dimensional crossover from 3D to quasi-2D and then 1D. In comparison, the red solid line denotes the one-step dimensional crossover from 3D to pure 2D studied in Ref. \cite{Zhou2010} with the red dotted curve being the pure 2D asymptotic behavior.}
      \label{figfx}
\end{figure}

We assume the lattice depths $V_1$ and $V_2$ in Eq.~(\ref{OL}) in unit of the recoil energy of $E_R$ are relatively large ($V_1\geq 5$, $V_2\geq 5$), so that the interband gap of $E_{\text{gap}}$ is bigger than the chemical potential of $\mu$, i. e. $E_{\text{gap}}\gg \mu$.  Meanwhile, because of the quantum tunneling, the overlap of the wave functions of
two consecutive wells are still sufficient to ensure full coherence even in the presence of disorder.
By this assumption~\cite{Orso2006,Hu2009,Hu2011,Zhou2010}, we restrict ourselves to the lowest band, where the
physics is governed by the ratio between the chemical potential $\mu$ and the bandwith of $4(J_1+J_2)$, where $J_1$ and $J_2$ are the tunneling rates
between neighboring wells. Generally speaking, for  $4(J_1+J_2)\gg \mu$,  the system retains an anisotropic 3D behavior, whereas for $4(J_1+J_2)\simeq \mu$, the system undergoes a dimensional crossover to a
1D regime.  In the limit of $4(J_1+J_2)\ll \mu $, the model system can be treated as 1D.
Following Refs.~\cite{Orso2006,Hu2009,Hu2011,Zhou2010}, we treat our model system within the tight-binding approximation as shown in Appendix \ref{AppendA}. The lowest Bloch band of the model system can be described in terms of the Wannier functions as
$\phi_{k_x}(x)\phi_{k_y}(y)$, with $\phi_{k_{x_i}}(x_i)=\sum_l{\rm e}^{{\rm i}ldk_{x_i}}w(x_i-ld)$. Here, $w(x_i)=\exp\left[-x_i^2/2\sigma_i^2\right]/\pi^{1/4}\sigma_i^{1/2}$  and $d/\sigma_i=\pi V_i^{1/4}\exp\left(-1/4\sqrt{V_i}\right)$ ($i=1,2$ and $x_1=x$, $x_2=y$). We remark that  this work is limited into a tight-binding approximation by neglecting beyond-lowest-Bloch-band transverse modes along the $x$ and $y$ directions. Further considering the effects of beyond-lowest-Bloch-band transverse modes on the dimensional crossover goes beyond the scope of this work.

Directly following Refs.~\cite{Hu2009,Zhou2010}, we expand the field operators in Hamiltonian (\ref{Model})
as $\hat{\Psi}(\bm {r})=\sum_{\bm {k}}\hat{a}_{\bm {k}}{\rm e}^{-{\rm i}k_z z}\phi_{k_x}(x)\phi_{k_y}(y)$ and obtain
\begin{eqnarray}
H-\mu N&=&\sum_{\bm{k}}\left(\varepsilon^0_{\bm{k}}-\mu\right)\hat{a}^\dagger_{\bm{k}}\hat{a}_{\bm{k}}+\frac{\tilde{g}}{2V}\sum_{\bm{k},\bm{q},
{\bm{k}}^\prime}\hat{a}^\dagger_{\bm{k}+\bm{q}}\hat{a}^\dagger_{{\bm{k}}^\prime-{\bm{q}}}\hat{a}_{{\bm{k}}^\prime}\hat{a}_{\bm{k}}\nonumber\\
&+&\sum_{{\bm{k}},{\bm{k}}^\prime}\hat{a}^\dagger_{\bm{k}}\hat{a}_{\bm{k}^\prime}V_{\bm{k}-\bm{k}^\prime},\label{EModel}
\end{eqnarray}
where
 \begin{align}
\varepsilon^0_{\bm{k}}=\frac{\hbar^2k_z^2}{2m}+2[J-J_1\cos k_x-J_2\cos k_y],\label{Single}
\end{align}
is the energy dispersion of the non-interacting system. Here, $J_1$ and $J_2$ are the tunneling rates along the $x$- and $y$-direction, respectively. Moreover, we have $J=J_1+J_2$, $V$ is the volume of the model system, and $\tilde{g}=gd^2/(2\pi\sigma_1\sigma_2)$ is the renormalized coupling constant. The $V_{\bm{k}}=1/V\int {\rm e}^{{\rm i}{\bm{k}}\cdot{\bm{r}}}V_{\text{ran}}{\rm d}\bm{r}$ in Eq.~(\ref{EModel}) is the Fourier transform of disorder potential.

We remark that in this work, we do not consider the effect of the confinement-induced resonance (CIR)~\cite{Peng2010,Zhang2011} on the coupling constant $\tilde{g}$. The basic physics of CIR can be understood in the language of Feshbach resonance~\cite{Bergeman2003}, where the scattering open channel and closed channels are, respectively, represented by the ground-state transverse mode and the other transverse modes along the tight-confinement dimensions. Within the tight-binding approximation assumed in this work, the ultracold atoms are frozen in the states of the lowest Bloch band and can not be excited into the other transverse modes. Thus the effect of CIR on $\tilde{g}$ can be safely ignored as the closed channels are absent~\cite{Peng2010,Zhang2011,Bergeman2003}.

Our subsequent calculations proceed in two steps. First, we calculate the ground state energy and quantum depletion. Previous studies~\cite{Hu2009,Zhou2010} have shown that the effects of disorder simply lead to trivial energy shifts in the ground state energy, and therefore, we shall ignore the disorder potential in this part of calculations and set $V_{\text{ran}}=0$. Second, we investigate how the dimensionality affects the superfluid density in the presence of the disorder potential $V_{\text{ran}}\neq 0$.

{\it Ground state energy and quantum depletion ---} For an optically-trapped Bose gas described by Hamiltonian (\ref{Model}), the ground state energy $E_g$ and quantum depletion
can be calculated via the single-particle Green function $G(\bm{k},\omega)$~\cite{Souza2021} as follows
\begin{equation}
\frac{E_g}{V}\!=\!\frac{\tilde{g}n^2}{2}\!\!+\!\!\lim\limits_{t\rightarrow0-}\!\!\frac{1}{(2\pi)^3 d^2}\!\!\int \!\!{\rm d}k_z{\rm d}^2\bm{k}\!\!\int\!\! \frac{{\rm d}\omega}{2\pi {\rm i}} {\rm e}^{-{\rm i}\omega t}\!E_{\bm{k}}G(\bm{k},\omega),\label{GSE}
\end{equation}
\begin{equation}
\frac{N-N_0}{N}\!=\!\lim\limits_{t\rightarrow0-}\frac{{\rm i}}{(2\pi)^4 d^2n}\int {\rm d}k_z{\rm d}^2\bm{k}\int {\rm d}\omega  {\rm e}^{-{\rm i}\omega t}G(\bm{k},\omega),\label{QD}
\end{equation}
with $E(\bm{k})$ being the excitation energy. In Eqs.~(\ref{GSE}) and (\ref{QD}), the $G(\bm{k},\omega)$ is the Fourier transformation of the Green function
\begin{equation}
G(\bm{k},t-t^\prime)=-{\rm i}\left \langle T\hat{a}_{\bm{k}}(t)\hat{a}_{\bm{k}}^\dagger(t^\prime)\right\rangle,
\end{equation}
in the Heisenberg representation, where $T$ denotes the chronological product.

By applying the Bogoliubov theory~\cite{Orso2006,Hu2009,Hu2011,Zhou2010,Faigle2021} to the Hamiltonian (\ref{Model}), we follow the standard procedures and obtain
\begin{align}
G(\bm{k},\omega)=\frac{\omega+\varepsilon^0_{\bm{k}}+\tilde{g}n}{\omega^2-E^2_{\bm{k}}+{\rm i}0}.\label{GF}
\end{align}
Here, $n_0$ is the condensate density, $E_{\bm{k}}=\sqrt{\varepsilon^0_{\bm{k}}(\varepsilon^0_{\bm{k}}+2\tilde{g}n)}$ and $\varepsilon^0_{\bm{k}}$ is defined in Eq.~(\ref{Single}).

By plugging Eq.~(\ref{GF}) into Eqs.~(\ref{GSE}) and (\ref{QD}), respectively, the ground state energy $E_g$ and quantum depletion $(N-N_0)/N$ are straightforwardly obtained (see the detailed derivations in Appendix \ref{AppendB})
\begin{align}
\frac{E_g}{V}=\frac{1}{2}\tilde{g}n^2-\frac{\sqrt{2m\tilde{g}n}\tilde{g}n}{4\pi\hbar d^2}f\left(\frac{2J}{\tilde{g}n}\right),\label{GSED}
\end{align}
and
\begin{equation}
\frac{N-N_0}{N}=\frac{1}{4\pi\hbar d^2}\sqrt{\frac{2m\tilde{g}}{n}}h\left(\frac{2J}{\tilde{g}n}\right).\label{QDD}
\end{equation}
In Eqs.~(\ref{GSED}) and (\ref{QDD}), the functions $f(s)$ and $h(s)$, respectively, are given by
\begin{align}
f(s)=\frac{\pi}{2}\int_{-\pi}^\pi \frac{{\rm d}^2{\bm{k}}}{(2\pi)^2}\frac{1}{\sqrt{s\gamma}}{}_2F_1\left[\frac{1}{2},\frac{3}{2},3,\frac{-2}{s\gamma}\right],\label{ffun}
\end{align}
and
\begin{equation}
h(s)=\int_{-\pi}^\pi \frac{{\rm d}^2{\bm{k}}}{(2\pi)^2}\int_0^\infty \frac{{\rm d}\eta}{\sqrt{\eta}}
\left[\frac{\eta+s\gamma+1}{\sqrt{(\eta+s\gamma)(\eta+s\gamma+2)}}-1\right].\label{hfun}
\end{equation}
In Eqs.~(\ref{ffun}) and (\ref{hfun}), the variable $s$ stands for $s=s_1+s_2=2(J_1+J_2)/\tilde{g}n$, which can be controlled by the strength of optical lattice in Eq.~(\ref{OL}), and $\gamma =1-(J_1/J)\cos k_x-(J_2/J)\cos k_y$. The function ${}_2F_1\left(a,b,c,d\right)$ in Eq.~(\ref{ffun}) is the hypergeometric function.

\begin{figure}
      \centering
      \includegraphics[width=0.5\textwidth]{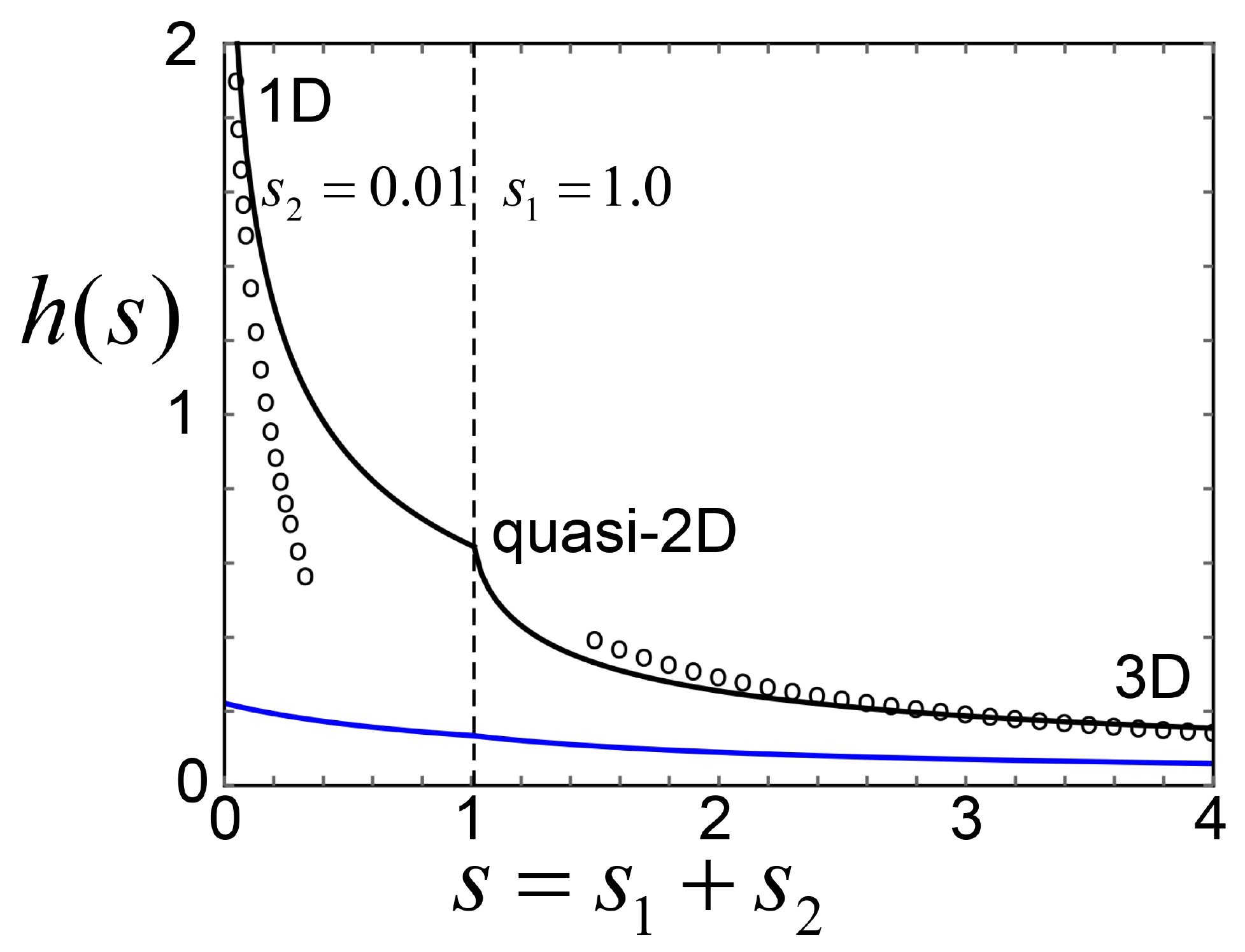}
      \caption{The behavior of $h(s)$ as the dimensionless tunneling rates of $s_{1,2}$ change independently. In the left side of the vertical dashed line, we fix $s_2=0.01$ and set $s_1=[0,1]$. In the right side, we fix $s_1=1$ and set $s_2=[0.01,3]$. The BEC behaves from 1D-like to quasi-2D like, and finally to 3D-like, as $s$ increases. The two black dotted lines denote the 1D and 3D asymptotic behaviors respectively. The blue curve describes the disorder-induced quantum depletion along the dimensional crossover, which is plotted by the functions $\frac{g^2_{\rm imp}n_{\rm imp}}{4\pi \tilde{g}^2n}\int_{-\pi}^{\pi}{\rm d}^2{\bm{k}}(2+s\gamma)^{-\frac{3}{2}}$ with $g^2_{\rm imp}n_{\rm imp}/({\tilde{g}^2n})=0.1$.}\label{fighx}
\end{figure}

Equations (\ref{GSED}) and (\ref{QDD}) are the key results of this work. In Figs.~\ref{figfx} and \ref{fighx}, we plot
$f(s)$ and $h(s)$, respectively. In the limit $s\rightarrow \infty$, the system is anisotropic 3D, whereas in the opposite limit $s\rightarrow 0$, the system is 1D. Thus, when continuously decreasing $s=2(J_1+J_2)/\tilde{g}n$ by enhancing the confinement, the system necessarily crossovers from the anisotropic 3D to 1D.We emphasize that the 3D-like gas here is referred as to an optically-trapped Bose gas in the tight-binding approximation, which is different from 3D Bose gas in the almost free space. However, from the theoretical angles, we can extend the parameter regimes from tight-binding-3D-gas to beyond-tight-binding-3D-gas, i.e. entering the parameter regime of $V_1< 5$, $V_2< 5$. In what follows, we are surprised to find that our analytical results can recover the Lee-Huang-Yang results obtained from the 3D free space as the surprising bonus of our analytical results. To induce the hierarchical dimensional crossover, we consider the following scheme for controlling the lattice depths, which consists of two stages: (i) we first fix the lattice strength $V_1=5$ and increase $V_2$ from the initial strength of $V_2=5$ to the final strength of $V_2=12$ (i.e., $s_2$ is decreased to almost zero), where the system is expected to crossover from the 3D to the quasi-2D; (ii) we fix $V_2=12$ and further increase the value of $V_1$ from the initial strength of $V_1=5$ to the final strength of $V_1=12$ until the value of $s=s_1+s_2$ is almost zero.

In the process (i), the behavior of the functions of $f(s)$ and $h(s)$ are shown by the solid curves in Figs.~\ref{figfx} and \ref{fighx}. Let us first check whether our analytical results in Eqs.~(\ref{GSED}) and (\ref{QDD}) in the limit $s\rightarrow \infty$ can recover the well-known 3D results of Bose gases. For $s\rightarrow \infty$, corresponding to the anisotropic 3D regime, we find  $f(s)\simeq1.43\sqrt{2/s}-32\sqrt{2}/(15\pi s)$ in Eq.~(\ref{ffun}) and $h(s)\simeq 8/(3\pi\sqrt{2}s)$, as denoted by the black circled curves in Figs.~\ref{figfx} and \ref{fighx}, respectively. Thus we exactly recover the 3D results of the quantum ground state energy and quantum depletion in Ref.~\cite{Zhou2010}. We note that our work is different from Ref.~\cite{Zhou2010}, where one adds a 1D optical lattice and increases the lattice depth to realize a purely 2D system. Instead, our scheme realizes the quasi-2D quantum system. To compare the two, we also plot the $f(s)$ associated with the case in Ref.~\cite{Zhou2010} [see red curves in Fig.~\ref{figfx}]. As clearly shown, our scheme realizes a quasi-2D (black curve), instead of a purely 2D, quantum system before it further crossovers to quasi-1D.

In the process (ii), we increase $V_1$ and fix the lattice depth $V_2$, where the system is expected to crossover from the quasi-2D to the quasi-1D and then to pure 1D. In particular,
we note that the function $f(s)$ shown in Fig.~\ref{figfx} exactly approaches $3\sqrt{2}/4$ in the limit $s\rightarrow 0$, corresponding to the Lieb-Liniger result of 1D Bose gas in Ref.~\cite{Orso2006}. For the quantum depletion shown in Fig.~\ref{fighx}, the function $h(s)$ diverges as $h(s)\simeq -\ln(1.35 s)/\sqrt{2}$. This signals
that in the absence of tunneling there is no real BEC, in agreement with the general theorems in one dimension.

Our results in Eqs.~(\ref{GSED}) and (\ref{QDD}) complement the descriptions of dimensional crossovers described in Refs.~\cite{Orso2006,Hu2009,Zhou2010,Hu2011}. We also note that the theoretical treatments beyond the Bogoliubov approximation are beyond the scope of this
work.

\begin{figure}
      \centering
      \includegraphics[width=0.5\textwidth]{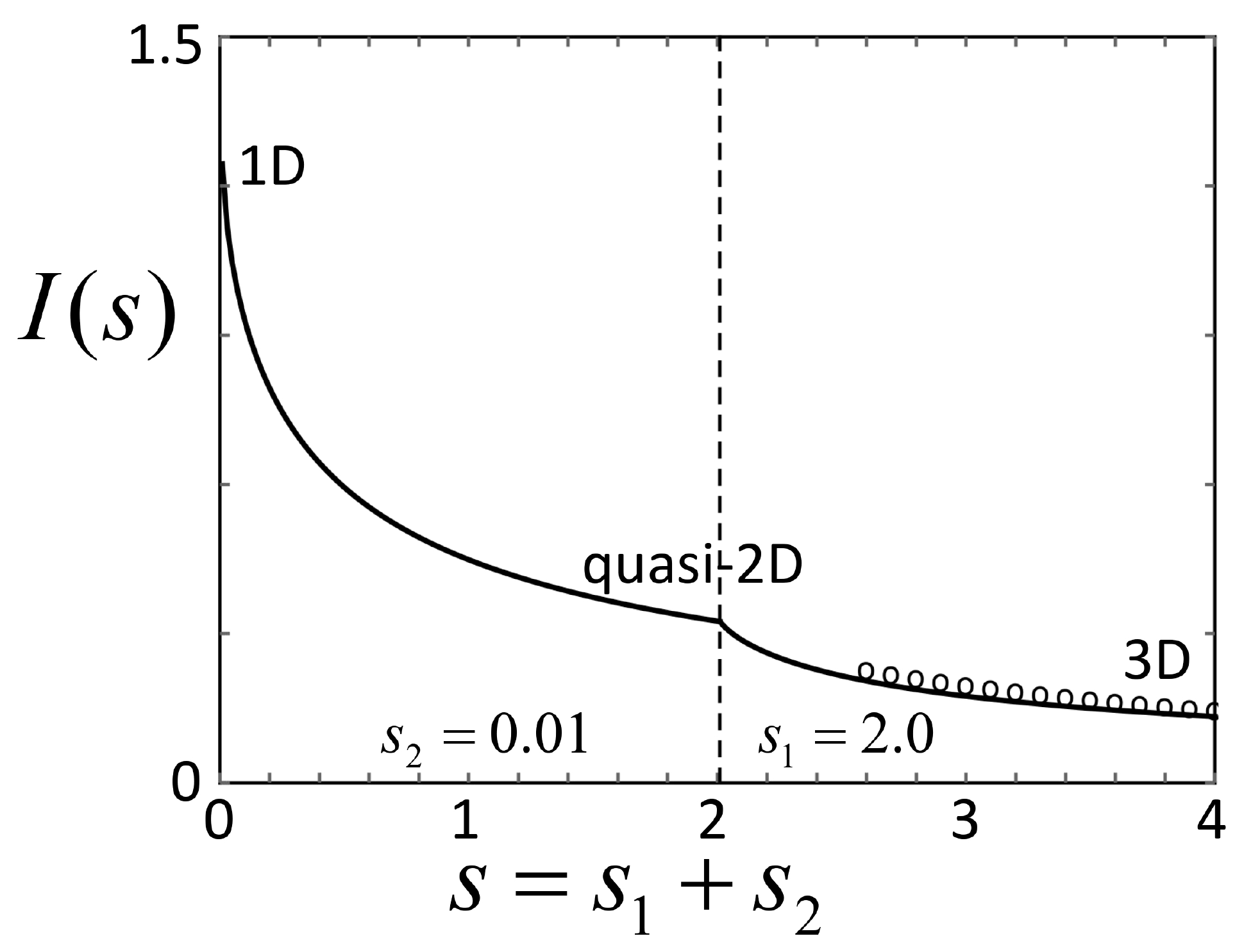}
      \caption{The behavior of $I(s)$ as the dimensionless tunneling rates of $s_{1,2}$ change independently. In the left side of the vertical dashed line, we fix $s_2=0.01$ and set $s_1=[0,2]$. In the right side, we fix $s_1=2$ and set $s_2=[0.01,2]$. The BEC behaves from 1D-like to quasi-2D like, and finally to 3D-like, as $s$ increases. The black dotted line in the right side denotes the 3D asymptotic behavior.}
      \label{figix}
\end{figure}

{\it Superfluid density ---} In the second part of this paper, we apply the linear response theory to investigate the effects of disorder on the
superfluid density of the BEC trapped in a 2D optical lattice. The superfluid density $\rho_s$ is determined by
the response of the momentum density to an externally imposed velocity field. We calculate $\rho_s$ based on the Bogoliubov approximation. Note that Ref.~\cite{Huang1992} pioneered in the study of the superfluid density of a 3D disordered Bose gas within the framework of Bogoliubov theory, which is consistent with the results obtained by the Beliaev-Popov diagrammatic technique \cite{Lopatin2002}. In the context of ultracold Bose gas, one of the authors in Refs.~\cite{Hu2009,Zhou2010} has investigated the disorder-induced superfluid density along the 3D-1D dimensional crossover using the Bogoliubov approximation.

In a disordered BEC, the static current-current response function consists of the low-frequency, long-wavelength longitudinal response $\chi_L\left(\bm{k}\right)$ and
the transverse response $\chi_T\left(\bm{k}\right)$, i.e., $\chi_{ij}\left(\bm{k}\right)=\frac{k_ik_j}{k^2} \chi_L\left(\bm{k}\right)+(\delta_{ij}-\frac{k_ik_j}{k^2})\chi_T\left(\bm{k}\right)$, see details of the definition of $\chi_{ij}$ in Refs.~\cite{Hu2009,Zhou2010}. The transverse response of a BEC is only due to the normal fluid, since the superfluid component can only participate in the irrotational flow.

For the disordered BEC trapped in a 2D optical
lattice described by Hamiltonian (\ref{Model}), where the rotational symmetry is broken, the response function along the unconfined $z$ direction is different from that in the confined $x$-$y$
plane. In the following, we assume a slow rotation
with respect to the $z$ axis and calculate the transverse
response function along the $z$ direction. We find
\begin{eqnarray}
\rho_n&=&\lim_{k\rightarrow 0}\chi_{T}\left(\bm{k}\right)=\chi_{zz}(0,0)=\frac{2n}{m}\sum\limits_p\frac{p_z^2\varepsilon^0_{\bm{k}}}
{E^4(\bm{k})}\langle|V_{\bm{k}}|^2\rangle\nonumber\\
&=&\frac{\tilde{R}\sqrt{2m\tilde{g}n}}{16\hbar d^2}I(s),\label{ifun}
\end{eqnarray}
where $\tilde{R}=n_{\rm imp}\tilde{b}^2/n\tilde{a}^2_{3D}$ and $I(s)$ with $s=s_1+s_2={2(J_1+J_2)}/{\tilde{g}{n}}$ is given by
\begin{align}
I(s)=\int_{-\pi}^{\pi}\frac{{\rm d}^2{\bm{k}}}{(2\pi)^2}\frac{2}{\sqrt{s\gamma+2}(s\gamma+1+\sqrt{s\gamma(s\gamma+2)})}.
\end{align}
Equation (\ref{ifun}) can be interpreted as the second-order term
in the perturbation expansion of the normal-fluid density in
terms of the weak disorder $V_{\bm{k}}$.

The result of Eq.~(\ref{ifun}) is plotted in Fig.~\ref{figix}. In the asymptotic 3D limit, one finds $I(s)\simeq 4\sqrt{2}/(3\pi s)$, corresponding to the dotted curve in Fig.~\ref{figix}. In
this case, Equation (\ref{ifun}) recovers the corresponding result of 3D Bose gases as in Ref.~\cite{Zhou2010}. Equation~(\ref{ifun}) presents another key result of this paper,
which provides an analytical expression for the normal fluid density
in a Bose fluid in an anisotropic two-dimensional optical lattice with the presence of weak
disorder.  The superfluid density $\rho_s=\rho-\rho_n$ is thus straightforwardly obtained.

{\it Discussion and conclusion ---} We justify the Bogoliubov approximation used in our calculations a {\it posteriori} by estimating the
quantum depletion~\cite{Orso2006}. The experimental work~\cite{Xu2006} by Ketterle's group has shown that the Bogoliubov theory
provides a semiquantitative description for an optically-trapped BEC even when the quantum depletions is in excess of $50\%$.
For a uniform BEC, the quantum depletion
is $(N-N_0)/N=8/3\sqrt{na_{3D}^2}$ and the Bogoliubov approximation is valid
provided $\sqrt{na_{3D}^2}$ is small. For an optically-trapped BEC, the quantum depletion is modified qualitatively as $(8m^*/3m)\sqrt{n\tilde{a}_{3D}/\pi}$ with $m^*$ being the effective mass, which remains small for typical experimental parameters as in Ref.~\cite{Orso2006}. For an optically-trapped Bose gas along the dimensional crossovers, we can estimate the quantum depletion $(N-N_0)/N$ with the help of Fig.~\ref{fighx}. Considering typical experiments in an optically-trapped BEC as in Ref.~\cite{Bloch2008}, the relevant parameters are: $n=3\times 10^{13}{\rm cm}^{-3}$, $d=430 {\rm nm}$, $a_{3D}=5.4 {\rm nm}$, and $d/\sigma_1\sim d/\sigma_2\sim 1$. The quantum depletion in Eq.~(\ref{QDD}) is thus evaluated as $(N-N_0)/N\sim 0.0036\times h(s)$, with $h(s)$ shown in Fig.~\ref{fighx}. It's clear that
the quantum depletion $(N-N_0)/N<20\%$, and therefore, the Bogoliubov approximation is valid in the sprit of Ref.~\cite{Xu2006}. Apart from the
phase fluctuations due to the tight confinement along $x$ and $y$ directions, the effect of the disorder potential can also
enhance quantum fluctuations and thus affect the Bogoliubov approximation. As such, we calculate the disorder-induced correction to the quantum depletion as
\begin{equation}
\frac{\Delta N^\prime}{N}=\frac{1}{4\pi\hbar d^2}\sqrt{\frac{2m\tilde{g}}{n}}\frac{g^2_{\rm imp}n_{\rm imp}}{4\pi \tilde{g}^2n}\int_{-\pi}^{\pi}{\rm d}^2{\bm{k}}(2+s\gamma)^{-\frac{3}{2}}.
\end{equation}
For the case of weak disorder of $g^2_{\rm imp}n_{\rm imp}/\tilde{g}^2n=0.1$ with $n_{\rm imp}$ being the impurity density, the quantum depletion due to the disorder along the dimensional crossover is shown by the blue curves in Fig.~\ref{fighx}. This result indicates the quantum depletion due to the disorder is small and the Bogoliubov approximation is still valid.

Summarizing, we have investigated a 3D disordered BEC trapped in an anisotropic 2D optical lattice characterized by the lattice depths of $V_1$ in $x$-direction and $V_2$ in $y$-direction, respectively. We have derived the analytical expressions of the ground-state energy, quantum depletion and superfluid density of the system. Our results show the hierarchical, 3D-quasi-2D-1D
crossovers in the behavior of quantum fluctuations and the superfluid density. The physics of the hierarchical dimensional crossover involves the
interplay of three quantities: the strength of the optical
lattice, the interaction between bosonic atoms, and the strength of disorder. All these quantities are experimentally
controllable using state-of-the-art technologies. In particular, the depth of an optical lattice can be
tuned from $0E_R$ to $32E_R$ almost at will~\cite{Bloch2008}. Therefore, the phenomena discussed
in this paper should be observable within the current experimental capabilities.
Observing this hierarchical dimensional effect directly
would present an important step in revealing the interplay
between dimensionality and quantum fluctuations in
quasi-low dimensions. The present work is based on the Bogoliubov theory.
Future studies along this direction include the treatment of the
system for the whole range of interatomic interaction
strength, from zero to infinity, as well as for arbitrarily strong
disorder.

We thank Chao Gao for stimulating discussions. This work was
supported by the Zhejiang Provincial Natural Science Foundation (Grant Nos. LZ21A040001 and LQ20A040004), the National Natural
Science Foundation of China (Nos. 12074344, and 12104407) and the key projects of the Natural Science Foundation of China (Grant No. 11835011).

{\it Note added.---} Before submitting our work, we notice that a similar work \cite{Yao2022} has studied the 2D-1D dimensional crossover. In contrast, our work
has focused on the gradual 3D-2D-1D dimensional crossover.

\appendix
\begin{widetext}
\section{Validity of tight-binding approximation}\label{AppendA}
In the tight-binding approximation~\cite{Bloch2008}, the tunnelling rates of $J_1$ and $J_2$ along the $x-$ and $y-$ directions are defined as
\begin{eqnarray}
J_1=\int_{-\infty}^{\infty}{\rm d}x w^*(x)\left(-\frac{\hbar^2}{2m}\frac{\partial^2}{\partial x^2}+V_1 \times E_R\sin^2(q_Bx)\right)w(x),\\
J_2=\int_{-\infty}^{\infty}{\rm d}y w^*(y)\left(-\frac{\hbar^2}{2m}\frac{\partial^2}{\partial y^2}+V_2 \times E_R\sin^2(q_By)\right)w(y),
\end{eqnarray}
with $w(x)$ and $w(y)$ being the Wannier functions in the $x-$ and $y-$ directions.  The analytic solutions for the Wannier functions can be obtained
by solving the 1D Mathieu problem as shown in Ref. \cite{Mathieu}. In such, the approximate analytic expressions of tunnelling rates $J_1$ and $J_2$ have been derived in Ref. \cite{TL}
\begin{eqnarray}
\frac{J_1}{E_R}=\frac{1}{4}\left(\frac{2}{\pi}\right)^{\frac{1}{2}}\left(\frac{V_1}{4}\right)^\frac{3}{4}2^5\exp\left(-2\sqrt{V_1}\right),\label {J1}\\
\frac{J_2}{E_R}=\frac{1}{4}\left(\frac{2}{\pi}\right)^{\frac{1}{2}}\left(\frac{V_2}{4}\right)^\frac{3}{4}2^5\exp\left(-2\sqrt{V_2}\right),\label{J2}
\end{eqnarray}
In order for analytical expressions of $J_1$ and $J_2$ to be valid, the considered energy band must be a slowly varying
function of the quasi-momentum. Hence, the potential depth s must be sufficiently large. Note that the tight-binding approximation is valid under the following conditions (i) Lattice depths od $V_1$ and $V_2$ in Eq.~(\ref{OL}) are relatively large ($V_1\geq 5$, $V_2\geq 5$) to make sure that the interband gap of $E_{\text{gap}}$ is bigger than the chemical potential of $\mu$, i. e. $E_{\text{gap}}\gg \mu$; (ii) The overlap of the wave functions of two consecutive wells are still sufficient to ensure full coherence because of the quantum tunneling.

Now we are ready to give the rough estimations of parameter regimes of the tight-binding approximation being valid. Here, we use the typically experimental parameters of an
optically-trapped Bose gas in Ref. \cite{Du2010}. The typical detailed parameters read as follows: the recoil energy is $E_R \approx h\times 3.33{\rm kHz}$ with $h$ being Plank constant and the chemical potential of gas is $\mu \approx \tilde{g}n\approx h \times 400 {\rm Hz}$. In the case of ($V_1\approx 5$, $V_2\approx 5$), we can estimate the parameters of $J_1/E_R\approx J_2/E_R \approx 0.09$ based on Eqs. (\ref{J1}) and (\ref{J2}). Then we can further estimate the dimensionless parameters used in the figures of this work as follows: $s=2(J_1+J_2)/\tilde{g}n=(2(J_1+J_2)/R_R) (E_R/\tilde{g}n)\approx 2(0.09+0.09)\times 10=3.6$, suggesting that our model system is 3D-like. Meanwhile, as shown in Ref. \cite{Du2010}, the optically-trapped Bose gas is entirely superfluid below the critical lattice height $V_{c}\approx 13 E_R$ corresponding to $J_1$ and $J_2$ being almost zero because of the exponential decrease in Eqs. (\ref{J1}) and (\ref{J2}). We conclude that the tight-binding approximation can be regarded to be valid under $5 < V_1 < 13 $ and $5< V_2 < 13 $, corresponding to $0<s < 4$ as shown in Figs. \ref{figfx} and \ref{fighx}.

\section{Detailed derivations of Eqs. (\ref{GSED}) and (\ref{QDD}).}\label{AppendB}

In this appendix, we  give the detailed derivations of Eqs. (\ref{GSED}) and (\ref{QDD}) .
The derivation of Eq. (\ref{GSED}) in the main text can be written as follows
\begin{align}
\frac{E_g}{V}=&\frac{\tilde{g}n^2}{2}+\lim\limits_{t\rightarrow0-}\frac{1}{(2\pi)^3 d^2}\int {\rm d}k_z{\rm d}^2{\bm{k}}\int \frac{{\rm d}\omega}{2\pi {\rm i}} {\rm e}^{-{\rm i}\omega t}E_{\bm{k}}G({\bm{k}},\omega)\notag\\
=&\frac{\tilde{g}n^2}{2}+\frac{1}{(2\pi)^3 d^2}\int {\rm d}k_z{\rm d}^2{\bm{k}}\int_C \frac{{\rm d}\omega}{2\pi {\rm i}} E_{\bm{k}}G({\bm{k}},\omega),
\end{align}
with
\begin{align}
   \int_C {\rm d}\omega G(\bm{k},\omega)=&\int_C {\rm d}\omega \frac{\omega+\varepsilon^0_{\bm{k}}+\tilde{g}n}{\omega^2-E^2_{\bm{k}}+{\rm i}0}\notag\\
   =&\int_C {\rm d}\omega \frac{\omega+\varepsilon^0_{\bm{k}}+\tilde{g}n}{2E_{\bm{k}}}(\frac{1}{\omega-E_{\bm{k}}+{\rm i}0}-\frac{1}{\omega+E_{\bm{k}}-{\rm i}0})\notag\\
   =&2\pi {\rm i}\lim\limits_{\omega\rightarrow-E_{\bm{k}}+{\rm i}0}\frac{\omega+\varepsilon^0_{\bm{k}}+\tilde{g}n}{2E_{\bm{k}}}(\omega+E_{\bm{k}}-{\rm i}0)
   \frac{-1}{\omega+E_{\bm{k}}-{\rm i}0}\notag\\
   =&-\pi {\rm i}\frac{\varepsilon^0_{\bm{k}}+\tilde{g}n-E_{\bm{k}}}{E_{\bm{k}}},
\end{align}
where we have used the residue theorem, and $C$ denotes the integration path around the upper half-plane.
Then we have
\begin{align}
   \frac{E_g}{V}=&\frac{\tilde{g}n^2}{2}-\frac{1}{(2\pi)^3d^22}\int_{-\pi}^\pi {\rm d}^2{\bm{k}}\int_{-\infty}^{\infty}{\rm d}k_z(\varepsilon^0_{\bm{k}}+\tilde{g}n-E_{\bm{k}})\notag\\
   =&\frac{\tilde{g}n^2}{2}-{\rm Int},
\end{align}
with
\begin{align}
{\rm Int}=&\frac{1}{(2\pi)^3d^22}\int_{-\pi}^\pi {\rm d}^2{\bm{k}}\int_{-\infty}^{\infty}{\rm d}k_z(\varepsilon^0_{\bm{k}}+\tilde{g}n-E_{\bm{k}})\notag\\
=&\frac{1}{(2\pi)^3d^22}\int_{-\pi}^\pi {\rm d}^2{\bm{k}}\int_{-\infty}^{\infty}{\rm d}k_z(\frac{\hbar^2k_z^2}{2m}+2J\gamma+\tilde{g}n-
\sqrt{(\frac{\hbar^2k_z^2}{2m}+2J\gamma)(\frac{\hbar^2k_z^2}{2m}+2J\gamma+2\tilde{g}n)})\notag\\
=&\frac{1}{(2\pi)^3d^22\hbar}\int_{-\pi}^\pi {\rm d}^2{\bm{k}}\int_{-\infty}^{\infty}{\rm d}p_z(\frac{p_z^2}{2m}+2J\gamma+\tilde{g}n-
\sqrt{(\frac{p_z^2}{2m}+2J\gamma)(\frac{p_z^2}{2m}+2J\gamma+2\tilde{g}n)})\notag\\
=&\frac{\sqrt{2m}2}{(2\pi)^3d^22\hbar}\int_{-\pi}^\pi {\rm d}^2{\bm{k}}\int_0^{\infty}{\rm d}p_z(p_z^2+2J\gamma+\tilde{g}n-
\sqrt{(p_z^2+2J\gamma)(p_z^2+2J\gamma+2\tilde{g}n)})\notag\\
=&\frac{\sqrt{2m}\tilde{g}n\sqrt{\tilde{g}n}}{(2\pi)^3d^2\hbar}\int_{-\pi}^\pi {\rm d}^2{\bm{k}}\int_0^{\infty}{\rm d}p_z(p_z^2+s\gamma+1-
\sqrt{(p_z^2+s\gamma)(p_z^2+s\gamma+2)}),
\end{align}
where $\gamma=1-\frac{J_1}{J}\cos(k_x)-\frac{J_2}{J}\cos(k_y)$ and $s=\frac{2J}{\tilde{g}n}$. We then let $p_z^2+s\gamma=\beta$,
\begin{align}
{\rm Int}=\frac{\sqrt{2m}\tilde{g}n\sqrt{\tilde{g}n}}{(2\pi)^3d^2\hbar}\int_{-\pi}^\pi {\rm d}^2{\bm{k}}\int_{s\gamma}^{\infty}\frac{{\rm d}\beta}{2\sqrt{\beta-s\gamma}}(\beta+1-\sqrt{\beta(\beta+2)}).
\end{align}
We then let $s\gamma/\beta=\tau$, and ${\rm d}\beta=-s\gamma {\rm d}\tau/\tau^2$,
\begin{align}
{\rm Int}=&\frac{\sqrt{2m}\tilde{g}n\sqrt{\tilde{g}n}}{(2\pi)^3d^2\hbar2}\int_{-\pi}^\pi {\rm d}^2{\bm{k}}\int_0^1
\frac{s\gamma}{\tau^2}{\rm d}\tau\frac{1}{\sqrt{\frac{s\gamma}{\tau}-s\gamma}}(\frac{s\gamma}{\tau}
+1-\sqrt{\frac{s\gamma}{\tau}(\frac{s\gamma}{\tau}+2)})\notag\\
=&\frac{\sqrt{2m}\tilde{g}n\sqrt{\tilde{g}n}}{(2\pi)^3d^2\hbar2}\int_{-\pi}^\pi {\rm d}^2{\bm{k}}\int_0^1{\rm d}\tau
\frac{\sqrt{s\gamma}}{\sqrt{\tau}}\frac{1}{\sqrt{1-\tau}}(\frac{s\gamma}{\tau^2}
+\frac{1}{\tau}-\frac{s\gamma}{\tau^2}\sqrt{1+\frac{2\tau}{s\gamma}})\notag\\
=&\frac{\sqrt{2m}\tilde{g}n\sqrt{\tilde{g}n}}{(2\pi)^3d^2\hbar2}\int_{-\pi}^\pi {\rm d}^2{\bm{k}}\frac{1}{\sqrt{s\gamma}}\int_0^1{\rm d}\tau
\frac{1}{\sqrt{\tau}}\frac{1}{\sqrt{1-\tau}}(\frac{(s\gamma)^2}{\tau^2}
+\frac{s\gamma}{\tau}-\frac{(s\gamma)^2}{\tau^2}\sqrt{1+\frac{2\tau}{s\gamma}}),
\end{align}
where the integration about $\tau$ can be written as the hypergeometric function
\begin{align}
\int_0^1{\rm d}\tau
\frac{1}{\sqrt{\tau}}\frac{1}{\sqrt{1-\tau}}(\frac{(s\gamma)^2}{\tau^2}
+\frac{s\gamma}{\tau}-\frac{(s\gamma)^2}{\tau^2}\sqrt{1+\frac{2\tau}{s\gamma}})
=\frac{\pi}{2}{}_2F_1[\frac{1}{2},\frac{3}{2},3,\frac{-2}{s\gamma}],
\end{align}
hence we obtain
\begin{align}
{\rm Int}=&\frac{\sqrt{2m}\tilde{g}n\sqrt{\tilde{g}n}}{(2\pi)^3d^2\hbar2}\int_{-\pi}^\pi {\rm d}^2{\bm{k}}\frac{1}{\sqrt{s\gamma}}\frac{\pi}{2}{}_2F_1[\frac{1}{2},\frac{3}{2},3,\frac{-2}{s\gamma}]\notag\\
=&\frac{\sqrt{2m\tilde{g}n}\tilde{g}n}{4\pi\hbar d^2}\frac{\pi}{2}\int_{-\pi}^\pi \frac{{\rm d}^2{\bm{k}}}{(2\pi)^2}\frac{1}{\sqrt{s\gamma}}{}_2F_1[\frac{1}{2},\frac{3}{2},3,\frac{-2}{s\gamma}]\notag\\
=&\frac{\sqrt{2m\tilde{g}n}\tilde{g}n}{4\pi\hbar d^2}f(s),
\end{align}
with
\begin{align}
f(s)=\frac{\pi}{2}\int_{-\pi}^\pi \frac{{\rm d}^2{\bm{k}}}{(2\pi)^2}\frac{1}{\sqrt{s\gamma}}{}_2F_1[\frac{1}{2},\frac{3}{2},3,\frac{-2}{s\gamma}].
\end{align}
Finally we get the Eqs. (\ref{GSED}) in the main text
\begin{align}
   \frac{E_g}{V}=&\frac{\tilde{g}n^2}{2}-{\rm Int}\notag\\
   =&\frac{\tilde{g}n^2}{2}-\frac{\sqrt{2m\tilde{g}n}\tilde{g}n}{4\pi\hbar d^2}f(s).
\end{align}

The derivation of Eqs. (\ref{QDD}) in the main text is as follows
\begin{align}
\frac{N-N_0}{N}=&\lim\limits_{t\rightarrow0-}\frac{{\rm i}}{(2\pi)^4 d^2n}\int {\rm d}k_z{\rm d}^2{\bm{k}}\int {\rm d}\omega  {\rm e}^{-{\rm i}\omega t}G(\bm{k},\omega)\notag\\
=&\frac{{\rm i}}{(2\pi)^4 d^2n}\int {\rm d}k_z{\rm d}^2{\bm{k}}\int_C {\rm d}\omega G(\bm{k},\omega)\notag\\
=&\frac{{\rm i}}{(2\pi)^4 d^2n}\int {\rm d}k_z{\rm d}^2{\bm{k}}(-\pi {\rm i})\frac{\varepsilon^0_{\bm{k}}+\tilde{g}n-E_{\bm{k}}}{E_{\bm{k}}}\notag\\
=&\frac{1}{2(2\pi)^3 d^2n}\int {\rm d}k_z{\rm d}^2{\bm{k}}\frac{\varepsilon^0_{\bm{k}}+\tilde{g}n-E_{\bm{k}}}{E_{\bm{k}}}\notag\\
=&\frac{1}{2(2\pi)^3 d^2n}\int_{-\pi}^{\pi}{\rm d}^2{\bm{k}}\int_{-\infty}^{\infty}{\rm d}k_z[\frac{\frac{\hbar^2k_z^2}{2m}+2J\gamma+\tilde{g}n}
{\sqrt{(\frac{\hbar^2k_z^2}{2m}+2J\gamma)(\frac{\hbar^2k_z^2}{2m}+2J\gamma+2\tilde{g}n)}}-1]\notag\\
=&\frac{1}{2(2\pi)^3\hbar d^2n}\int_{-\pi}^{\pi}{\rm d}^2{\bm{k}}\int_{-\infty}^{\infty}{\rm d}p_z[\frac{\frac{p_z^2}{2m}+2J\gamma+\tilde{g}n}
{\sqrt{(\frac{p_z^2}{2m}+2J\gamma)(\frac{p_z^2}{2m}+2J\gamma+2\tilde{g}n)}}-1]\notag\\
=&\frac{\sqrt{2m}}{2(2\pi)^3\hbar d^2n}\int_{-\pi}^{\pi}{\rm d}^2{\bm{k}}\int_{-\infty}^{\infty}{\rm d}p_z[\frac{p_z^2+2J\gamma+\tilde{g}n}
{\sqrt{(p_z^2+2J\gamma)(p_z^2+2J\gamma+2\tilde{g}n)}}-1]\notag\\
=&\frac{\sqrt{2m}}{2(2\pi)^3\hbar d^2n}\int_{-\pi}^{\pi}{\rm d}^2{\bm{k}}2\int_0^\infty \frac{{\rm d}\eta}{2\sqrt{\eta}}[\frac{\eta+2J\gamma+\tilde{g}n}
{\sqrt{(\eta+2J\gamma)(\eta+2J\gamma+2\tilde{g}n)}}-1]\notag\\
=&\frac{\sqrt{2m\tilde{g}n}}{2(2\pi)^3\hbar d^2n}\int_{-\pi}^{\pi}{\rm d}^2{\bm{k}}\int_0^\infty \frac{{\rm d}\eta}{\sqrt{\eta}}[\frac{\eta+s\gamma+1}
{\sqrt{(\eta+s\gamma)(\eta+s\gamma+2)}}-1]\notag\\
=&\frac{1}{4\pi\hbar d^2}\sqrt{\frac{2m\tilde{g}}{n}}\int_{-\pi}^{\pi}\frac{{\rm d}^2{\bm{k}}}{(2\pi)^2}\int_0^\infty \frac{{\rm d}\eta}{\sqrt{\eta}}[\frac{\eta+s\gamma+1}
{\sqrt{(\eta+s\gamma)(\eta+s\gamma+2)}}-1]\notag\\
=&\frac{1}{4\pi\hbar d^2}\sqrt{\frac{2m\tilde{g}}{n}}h(s),
\end{align}
with
\begin{align}
h(s)=\int_{-\pi}^{\pi}\frac{{\rm d}^2{\bm{k}}}{(2\pi)^2}\int_0^\infty \frac{{\rm d}\eta}{\sqrt{\eta}}[\frac{\eta+s\gamma+1}
{\sqrt{(\eta+s\gamma)(\eta+s\gamma+2)}}-1].
\end{align}
\end{widetext}
\bibliography{Reference}

\begin{thebibliography}{32}%
\makeatletter
\providecommand \@ifxundefined [1]{%
 \@ifx{#1\undefined}
}%
\providecommand \@ifnum [1]{%
 \ifnum #1\expandafter \@firstoftwo
 \else \expandafter \@secondoftwo
 \fi
}%
\providecommand \@ifx [1]{%
 \ifx #1\expandafter \@firstoftwo
 \else \expandafter \@secondoftwo
 \fi
}%
\providecommand \natexlab [1]{#1}%
\providecommand \enquote  [1]{``#1''}%
\providecommand \bibnamefont  [1]{#1}%
\providecommand \bibfnamefont [1]{#1}%
\providecommand \citenamefont [1]{#1}%
\providecommand \href@noop [0]{\@secondoftwo}%
\providecommand \href [0]{\begingroup \@sanitize@url \@href}%
\providecommand \@href[1]{\@@startlink{#1}\@@href}%
\providecommand \@@href[1]{\endgroup#1\@@endlink}%
\providecommand \@sanitize@url [0]{\catcode `\\12\catcode `\$12\catcode
  `\&12\catcode `\#12\catcode `\^12\catcode `\_12\catcode `\%12\relax}%
\providecommand \@@startlink[1]{}%
\providecommand \@@endlink[0]{}%
\providecommand \url  [0]{\begingroup\@sanitize@url \@url }%
\providecommand \@url [1]{\endgroup\@href {#1}{\urlprefix }}%
\providecommand \urlprefix  [0]{URL }%
\providecommand \Eprint [0]{\href }%
\providecommand \doibase [0]{http://dx.doi.org/}%
\providecommand \selectlanguage [0]{\@gobble}%
\providecommand \bibinfo  [0]{\@secondoftwo}%
\providecommand \bibfield  [0]{\@secondoftwo}%
\providecommand \translation [1]{[#1]}%
\providecommand \BibitemOpen [0]{}%
\providecommand \bibitemStop [0]{}%
\providecommand \bibitemNoStop [0]{.\EOS\space}%
\providecommand \EOS [0]{\spacefactor3000\relax}%
\providecommand \BibitemShut  [1]{\csname bibitem#1\endcsname}%
\let\auto@bib@innerbib\@empty
\bibitem [{\citenamefont {Lee}\ \emph {et~al.}(2006)\citenamefont {Lee},
  \citenamefont {Nagaosa},\ and\ \citenamefont {Wen}}]{Lee2006}%
  \BibitemOpen
  \bibfield  {author} {\bibinfo {author} {\bibfnamefont {Patrick~A.}\
  \bibnamefont {Lee}}, \bibinfo {author} {\bibfnamefont {Naoto}\ \bibnamefont
  {Nagaosa}}, \ and\ \bibinfo {author} {\bibfnamefont {Xiao-Gang}\ \bibnamefont
  {Wen}},\ }\bibfield  {title} {\enquote {\bibinfo {title} {Doping a mott
  insulator: Physics of high-temperature superconductivity},}\ }\href {\doibase
  10.1103/RevModPhys.78.17} {\bibfield  {journal} {\bibinfo  {journal} {Rev.
  Mod. Phys.}\ }\textbf {\bibinfo {volume} {78}},\ \bibinfo {pages} {17--85}
  (\bibinfo {year} {2006})}\BibitemShut {NoStop}%
\bibitem [{\citenamefont {Cao}\ \emph {et~al.}(2018{\natexlab{a}})\citenamefont
  {Cao}, \citenamefont {Fatemi}, \citenamefont {Demir}, \citenamefont {Fang},
  \citenamefont {Tomarken}, \citenamefont {Luo}, \citenamefont
  {Sanchez-Yamagishi}, \citenamefont {Watanabe}, \citenamefont {Taniguchi},
  \citenamefont {Kaxiras}, \citenamefont {Ashoori},\ and\ \citenamefont
  {Jarillo-Herrero}}]{Cao2018a}%
  \BibitemOpen
  \bibfield  {author} {\bibinfo {author} {\bibfnamefont {Yuan}\ \bibnamefont
  {Cao}}, \bibinfo {author} {\bibfnamefont {Valla}\ \bibnamefont {Fatemi}},
  \bibinfo {author} {\bibfnamefont {Ahmet}\ \bibnamefont {Demir}}, \bibinfo
  {author} {\bibfnamefont {Shiang}\ \bibnamefont {Fang}}, \bibinfo {author}
  {\bibfnamefont {Spencer~L.}\ \bibnamefont {Tomarken}}, \bibinfo {author}
  {\bibfnamefont {Jason~Y.}\ \bibnamefont {Luo}}, \bibinfo {author}
  {\bibfnamefont {Javier~D.}\ \bibnamefont {Sanchez-Yamagishi}}, \bibinfo
  {author} {\bibfnamefont {Kenji}\ \bibnamefont {Watanabe}}, \bibinfo {author}
  {\bibfnamefont {Takashi}\ \bibnamefont {Taniguchi}}, \bibinfo {author}
  {\bibfnamefont {Efthimios}\ \bibnamefont {Kaxiras}}, \bibinfo {author}
  {\bibfnamefont {Ray~C.}\ \bibnamefont {Ashoori}}, \ and\ \bibinfo {author}
  {\bibfnamefont {Pablo}\ \bibnamefont {Jarillo-Herrero}},\ }\bibfield  {title}
  {\enquote {\bibinfo {title} {Correlated insulator behaviour at half-filling
  in magic-angle graphene superlattices},}\ }\href {\doibase
  10.1038/nature26154} {\bibfield  {journal} {\bibinfo  {journal} {Nature}\
  }\textbf {\bibinfo {volume} {556}},\ \bibinfo {pages} {80--84} (\bibinfo
  {year} {2018}{\natexlab{a}})}\BibitemShut {NoStop}%
\bibitem [{\citenamefont {Cao}\ \emph {et~al.}(2018{\natexlab{b}})\citenamefont
  {Cao}, \citenamefont {Fatemi}, \citenamefont {Demir}, \citenamefont {Fang},
  \citenamefont {Tomarken}, \citenamefont {Luo}, \citenamefont
  {Sanchez-Yamagishi}, \citenamefont {Watanabe}, \citenamefont {Taniguchi},
  \citenamefont {Kaxiras}, \citenamefont {Ashoori},\ and\ \citenamefont
  {Jarillo-Herrero}}]{Cao2018b}%
  \BibitemOpen
  \bibfield  {author} {\bibinfo {author} {\bibfnamefont {Yuan}\ \bibnamefont
  {Cao}}, \bibinfo {author} {\bibfnamefont {Valla}\ \bibnamefont {Fatemi}},
  \bibinfo {author} {\bibfnamefont {Ahmet}\ \bibnamefont {Demir}}, \bibinfo
  {author} {\bibfnamefont {Shiang}\ \bibnamefont {Fang}}, \bibinfo {author}
  {\bibfnamefont {Spencer~L.}\ \bibnamefont {Tomarken}}, \bibinfo {author}
  {\bibfnamefont {Jason~Y.}\ \bibnamefont {Luo}}, \bibinfo {author}
  {\bibfnamefont {Javier~D.}\ \bibnamefont {Sanchez-Yamagishi}}, \bibinfo
  {author} {\bibfnamefont {Kenji}\ \bibnamefont {Watanabe}}, \bibinfo {author}
  {\bibfnamefont {Takashi}\ \bibnamefont {Taniguchi}}, \bibinfo {author}
  {\bibfnamefont {Efthimios}\ \bibnamefont {Kaxiras}}, \bibinfo {author}
  {\bibfnamefont {Ray~C.}\ \bibnamefont {Ashoori}}, \ and\ \bibinfo {author}
  {\bibfnamefont {Pablo}\ \bibnamefont {Jarillo-Herrero}},\ }\bibfield  {title}
  {\enquote {\bibinfo {title} {Correlated insulator behaviour at half-filling
  in magic-angle graphene superlattices},}\ }\href {\doibase
  10.1038/nature26154} {\bibfield  {journal} {\bibinfo  {journal} {Nature}\
  }\textbf {\bibinfo {volume} {556}},\ \bibinfo {pages} {80--84} (\bibinfo
  {year} {2018}{\natexlab{b}})}\BibitemShut {NoStop}%
\bibitem [{\citenamefont {Tarnopolsky}\ \emph {et~al.}(2019)\citenamefont
  {Tarnopolsky}, \citenamefont {Kruchkov},\ and\ \citenamefont
  {Vishwanath}}]{Tarnopolsky2019}%
  \BibitemOpen
  \bibfield  {author} {\bibinfo {author} {\bibfnamefont {Grigory}\ \bibnamefont
  {Tarnopolsky}}, \bibinfo {author} {\bibfnamefont {Alex~Jura}\ \bibnamefont
  {Kruchkov}}, \ and\ \bibinfo {author} {\bibfnamefont {Ashvin}\ \bibnamefont
  {Vishwanath}},\ }\bibfield  {title} {\enquote {\bibinfo {title} {Origin of
  magic angles in twisted bilayer graphene},}\ }\href {\doibase
  10.1103/PhysRevLett.122.106405} {\bibfield  {journal} {\bibinfo  {journal}
  {Phys. Rev. Lett.}\ }\textbf {\bibinfo {volume} {122}},\ \bibinfo {pages}
  {106405} (\bibinfo {year} {2019})}\BibitemShut {NoStop}%
\bibitem [{\citenamefont {Haldane}(1981)}]{Haldane1981}%
  \BibitemOpen
  \bibfield  {author} {\bibinfo {author} {\bibfnamefont {F.~D.~M.}\
  \bibnamefont {Haldane}},\ }\bibfield  {title} {\enquote {\bibinfo {title}
  {Effective harmonic-fluid approach to low-energy properties of
  one-dimensional quantum fluids},}\ }\href {\doibase
  10.1103/PhysRevLett.47.1840} {\bibfield  {journal} {\bibinfo  {journal}
  {Phys. Rev. Lett.}\ }\textbf {\bibinfo {volume} {47}},\ \bibinfo {pages}
  {1840--1843} (\bibinfo {year} {1981})}\BibitemShut {NoStop}%
\bibitem [{\citenamefont {Bloch}\ \emph {et~al.}(2008)\citenamefont {Bloch},
  \citenamefont {Dalibard},\ and\ \citenamefont {Zwerger}}]{Bloch2008}%
  \BibitemOpen
  \bibfield  {author} {\bibinfo {author} {\bibfnamefont {Immanuel}\
  \bibnamefont {Bloch}}, \bibinfo {author} {\bibfnamefont {Jean}\ \bibnamefont
  {Dalibard}}, \ and\ \bibinfo {author} {\bibfnamefont {Wilhelm}\ \bibnamefont
  {Zwerger}},\ }\bibfield  {title} {\enquote {\bibinfo {title} {Many-body
  physics with ultracold gases},}\ }\href {\doibase 10.1103/RevModPhys.80.885}
  {\bibfield  {journal} {\bibinfo  {journal} {Rev. Mod. Phys.}\ }\textbf
  {\bibinfo {volume} {80}},\ \bibinfo {pages} {885--964} (\bibinfo {year}
  {2008})}\BibitemShut {NoStop}%
\bibitem [{\citenamefont {Paredes}\ \emph {et~al.}(2004)\citenamefont
  {Paredes}, \citenamefont {Widera}, \citenamefont {Murg}, \citenamefont
  {Mandel}, \citenamefont {F{\"o}lling}, \citenamefont {Cirac}, \citenamefont
  {Shlyapnikov}, \citenamefont {H{\"a}nsch},\ and\ \citenamefont
  {Bloch}}]{Paredes2004}%
  \BibitemOpen
  \bibfield  {author} {\bibinfo {author} {\bibfnamefont {Bel{\'e}n}\
  \bibnamefont {Paredes}}, \bibinfo {author} {\bibfnamefont {Artur}\
  \bibnamefont {Widera}}, \bibinfo {author} {\bibfnamefont {Valentin}\
  \bibnamefont {Murg}}, \bibinfo {author} {\bibfnamefont {Olaf}\ \bibnamefont
  {Mandel}}, \bibinfo {author} {\bibfnamefont {Simon}\ \bibnamefont
  {F{\"o}lling}}, \bibinfo {author} {\bibfnamefont {Ignacio}\ \bibnamefont
  {Cirac}}, \bibinfo {author} {\bibfnamefont {Gora~V.}\ \bibnamefont
  {Shlyapnikov}}, \bibinfo {author} {\bibfnamefont {Theodor~W.}\ \bibnamefont
  {H{\"a}nsch}}, \ and\ \bibinfo {author} {\bibfnamefont {Immanuel}\
  \bibnamefont {Bloch}},\ }\bibfield  {title} {\enquote {\bibinfo {title}
  {Tonks--girardeau gas of ultracold atoms in an optical lattice},}\ }\href
  {\doibase 10.1038/nature02530} {\bibfield  {journal} {\bibinfo  {journal}
  {Nature}\ }\textbf {\bibinfo {volume} {429}},\ \bibinfo {pages} {277--281}
  (\bibinfo {year} {2004})}\BibitemShut {NoStop}%
\bibitem [{\citenamefont {Peppler}\ \emph {et~al.}(2018)\citenamefont
  {Peppler}, \citenamefont {Dyke}, \citenamefont {Zamorano}, \citenamefont
  {Herrera}, \citenamefont {Hoinka},\ and\ \citenamefont {Vale}}]{Peppler2018}%
  \BibitemOpen
  \bibfield  {author} {\bibinfo {author} {\bibfnamefont {T.}~\bibnamefont
  {Peppler}}, \bibinfo {author} {\bibfnamefont {P.}~\bibnamefont {Dyke}},
  \bibinfo {author} {\bibfnamefont {M.}~\bibnamefont {Zamorano}}, \bibinfo
  {author} {\bibfnamefont {I.}~\bibnamefont {Herrera}}, \bibinfo {author}
  {\bibfnamefont {S.}~\bibnamefont {Hoinka}}, \ and\ \bibinfo {author}
  {\bibfnamefont {C.~J.}\ \bibnamefont {Vale}},\ }\bibfield  {title} {\enquote
  {\bibinfo {title} {Quantum anomaly and 2d-3d crossover in strongly
  interacting fermi gases},}\ }\href {\doibase 10.1103/PhysRevLett.121.120402}
  {\bibfield  {journal} {\bibinfo  {journal} {Phys. Rev. Lett.}\ }\textbf
  {\bibinfo {volume} {121}},\ \bibinfo {pages} {120402} (\bibinfo {year}
  {2018})}\BibitemShut {NoStop}%
\bibitem [{\citenamefont {Holten}\ \emph {et~al.}(2018)\citenamefont {Holten},
  \citenamefont {Bayha}, \citenamefont {Klein}, \citenamefont {Murthy},
  \citenamefont {Preiss},\ and\ \citenamefont {Jochim}}]{Holten2018}%
  \BibitemOpen
  \bibfield  {author} {\bibinfo {author} {\bibfnamefont {M.}~\bibnamefont
  {Holten}}, \bibinfo {author} {\bibfnamefont {L.}~\bibnamefont {Bayha}},
  \bibinfo {author} {\bibfnamefont {A.~C.}\ \bibnamefont {Klein}}, \bibinfo
  {author} {\bibfnamefont {P.~A.}\ \bibnamefont {Murthy}}, \bibinfo {author}
  {\bibfnamefont {P.~M.}\ \bibnamefont {Preiss}}, \ and\ \bibinfo {author}
  {\bibfnamefont {S.}~\bibnamefont {Jochim}},\ }\bibfield  {title} {\enquote
  {\bibinfo {title} {Anomalous breaking of scale invariance in a
  two-dimensional fermi gas},}\ }\href {\doibase
  10.1103/PhysRevLett.121.120401} {\bibfield  {journal} {\bibinfo  {journal}
  {Phys. Rev. Lett.}\ }\textbf {\bibinfo {volume} {121}},\ \bibinfo {pages}
  {120401} (\bibinfo {year} {2018})}\BibitemShut {NoStop}%
\bibitem [{\citenamefont {Orso}\ \emph {et~al.}(2006)\citenamefont {Orso},
  \citenamefont {Menotti},\ and\ \citenamefont {Stringari}}]{Orso2006}%
  \BibitemOpen
  \bibfield  {author} {\bibinfo {author} {\bibfnamefont {G.}~\bibnamefont
  {Orso}}, \bibinfo {author} {\bibfnamefont {C.}~\bibnamefont {Menotti}}, \
  and\ \bibinfo {author} {\bibfnamefont {S.}~\bibnamefont {Stringari}},\
  }\bibfield  {title} {\enquote {\bibinfo {title} {Quantum fluctuations and
  collective oscillations of a bose-einstein condensate in a 2d optical
  lattice},}\ }\href {\doibase 10.1103/PhysRevLett.97.190408} {\bibfield
  {journal} {\bibinfo  {journal} {Phys. Rev. Lett.}\ }\textbf {\bibinfo
  {volume} {97}},\ \bibinfo {pages} {190408} (\bibinfo {year}
  {2006})}\BibitemShut {NoStop}%
\bibitem [{\citenamefont {Hu}\ \emph {et~al.}(2009)\citenamefont {Hu},
  \citenamefont {Liang},\ and\ \citenamefont {Hu}}]{Hu2009}%
  \BibitemOpen
  \bibfield  {author} {\bibinfo {author} {\bibfnamefont {Ying}\ \bibnamefont
  {Hu}}, \bibinfo {author} {\bibfnamefont {Zhaoxin}\ \bibnamefont {Liang}}, \
  and\ \bibinfo {author} {\bibfnamefont {Bambi}\ \bibnamefont {Hu}},\
  }\bibfield  {title} {\enquote {\bibinfo {title} {Effects of disorder on
  quantum fluctuations and superfluid density of a bose-einstein condensate in
  a two-dimensional optical lattice},}\ }\href {\doibase
  10.1103/PhysRevA.80.043629} {\bibfield  {journal} {\bibinfo  {journal} {Phys.
  Rev. A}\ }\textbf {\bibinfo {volume} {80}},\ \bibinfo {pages} {043629}
  (\bibinfo {year} {2009})}\BibitemShut {NoStop}%
\bibitem [{\citenamefont {Hu}\ and\ \citenamefont {Liang}(2011)}]{Hu2011}%
  \BibitemOpen
  \bibfield  {author} {\bibinfo {author} {\bibfnamefont {Ying}\ \bibnamefont
  {Hu}}\ and\ \bibinfo {author} {\bibfnamefont {Zhaoxin}\ \bibnamefont
  {Liang}},\ }\bibfield  {title} {\enquote {\bibinfo {title} {Visualization of
  dimensional effects in collective excitations of optically trapped
  quasi-two-dimensional bose gases},}\ }\href {\doibase
  10.1103/PhysRevLett.107.110401} {\bibfield  {journal} {\bibinfo  {journal}
  {Phys. Rev. Lett.}\ }\textbf {\bibinfo {volume} {107}},\ \bibinfo {pages}
  {110401} (\bibinfo {year} {2011})}\BibitemShut {NoStop}%
\bibitem [{\citenamefont {Zhou}\ \emph {et~al.}(2010)\citenamefont {Zhou},
  \citenamefont {Hu}, \citenamefont {Liang},\ and\ \citenamefont
  {Zhang}}]{Zhou2010}%
  \BibitemOpen
  \bibfield  {author} {\bibinfo {author} {\bibfnamefont {Kezhao}\ \bibnamefont
  {Zhou}}, \bibinfo {author} {\bibfnamefont {Ying}\ \bibnamefont {Hu}},
  \bibinfo {author} {\bibfnamefont {Zhaoxin}\ \bibnamefont {Liang}}, \ and\
  \bibinfo {author} {\bibfnamefont {Zhidong}\ \bibnamefont {Zhang}},\
  }\bibfield  {title} {\enquote {\bibinfo {title} {Optically trapped
  quasi-two-dimensional bose gases in a random environment: Quantum
  fluctuations and superfluid density},}\ }\href {\doibase
  10.1103/PhysRevA.82.043609} {\bibfield  {journal} {\bibinfo  {journal} {Phys.
  Rev. A}\ }\textbf {\bibinfo {volume} {82}},\ \bibinfo {pages} {043609}
  (\bibinfo {year} {2010})}\BibitemShut {NoStop}%
\bibitem [{\citenamefont {Faigle-Cedzich}\ \emph {et~al.}(2021)\citenamefont
  {Faigle-Cedzich}, \citenamefont {Pawlowski},\ and\ \citenamefont
  {Wetterich}}]{Faigle2021}%
  \BibitemOpen
  \bibfield  {author} {\bibinfo {author} {\bibfnamefont {Bruno~M.}\
  \bibnamefont {Faigle-Cedzich}}, \bibinfo {author} {\bibfnamefont {Jan~M.}\
  \bibnamefont {Pawlowski}}, \ and\ \bibinfo {author} {\bibfnamefont
  {Christof}\ \bibnamefont {Wetterich}},\ }\bibfield  {title} {\enquote
  {\bibinfo {title} {Dimensional crossover in ultracold fermi gases from
  functional renormalization},}\ }\href {\doibase 10.1103/PhysRevA.103.033320}
  {\bibfield  {journal} {\bibinfo  {journal} {Phys. Rev. A}\ }\textbf {\bibinfo
  {volume} {103}},\ \bibinfo {pages} {033320} (\bibinfo {year}
  {2021})}\BibitemShut {NoStop}%
\bibitem [{\citenamefont {Hu}\ \emph {et~al.}(2019)\citenamefont {Hu},
  \citenamefont {Mulkerin}, \citenamefont {Toniolo}, \citenamefont {He},\ and\
  \citenamefont {Liu}}]{Hu2019}%
  \BibitemOpen
  \bibfield  {author} {\bibinfo {author} {\bibfnamefont {Hui}\ \bibnamefont
  {Hu}}, \bibinfo {author} {\bibfnamefont {Brendan~C.}\ \bibnamefont
  {Mulkerin}}, \bibinfo {author} {\bibfnamefont {Umberto}\ \bibnamefont
  {Toniolo}}, \bibinfo {author} {\bibfnamefont {Lianyi}\ \bibnamefont {He}}, \
  and\ \bibinfo {author} {\bibfnamefont {Xia-Ji}\ \bibnamefont {Liu}},\
  }\bibfield  {title} {\enquote {\bibinfo {title} {Reduced quantum anomaly in a
  quasi-two-dimensional fermi superfluid: Significance of the
  confinement-induced effective range of interactions},}\ }\href {\doibase
  10.1103/PhysRevLett.122.070401} {\bibfield  {journal} {\bibinfo  {journal}
  {Phys. Rev. Lett.}\ }\textbf {\bibinfo {volume} {122}},\ \bibinfo {pages}
  {070401} (\bibinfo {year} {2019})}\BibitemShut {NoStop}%
\bibitem [{\citenamefont {Yin}\ \emph {et~al.}(2020)\citenamefont {Yin},
  \citenamefont {Hu},\ and\ \citenamefont {Liu}}]{Yin2020}%
  \BibitemOpen
  \bibfield  {author} {\bibinfo {author} {\bibfnamefont {X.~Y.}\ \bibnamefont
  {Yin}}, \bibinfo {author} {\bibfnamefont {Hui}\ \bibnamefont {Hu}}, \ and\
  \bibinfo {author} {\bibfnamefont {Xia-Ji}\ \bibnamefont {Liu}},\ }\bibfield
  {title} {\enquote {\bibinfo {title} {Few-body perspective of a quantum
  anomaly in two-dimensional fermi gases},}\ }\href {\doibase
  10.1103/PhysRevLett.124.013401} {\bibfield  {journal} {\bibinfo  {journal}
  {Phys. Rev. Lett.}\ }\textbf {\bibinfo {volume} {124}},\ \bibinfo {pages}
  {013401} (\bibinfo {year} {2020})}\BibitemShut {NoStop}%
\bibitem [{\citenamefont {Yao}\ \emph {et~al.}()\citenamefont {Yao},
  \citenamefont {Pizzino},\ and\ \citenamefont {Giamarchi}}]{Yao2022}%
  \BibitemOpen
  \bibfield  {author} {\bibinfo {author} {\bibfnamefont {Hepeng}\ \bibnamefont
  {Yao}}, \bibinfo {author} {\bibfnamefont {Lorenzo}\ \bibnamefont {Pizzino}},
  \ and\ \bibinfo {author} {\bibfnamefont {Thierry}\ \bibnamefont
  {Giamarchi}},\ }\bibfield  {title} {\enquote {\bibinfo {title}
  {Strongly-interacting bosons at 2d-1d dimensional crossover},}\ }\href
  {https://arxiv.org/abs/2204.02240} {\bibinfo  {journal} {arXiv:2204.02240v1}\
  }\BibitemShut {NoStop}%
\bibitem [{\citenamefont {White}\ \emph {et~al.}(2009)\citenamefont {White},
  \citenamefont {Pasienski}, \citenamefont {McKay}, \citenamefont {Zhou},
  \citenamefont {Ceperley},\ and\ \citenamefont {DeMarco}}]{White2009}%
  \BibitemOpen
\bibfield  {journal} {  }\bibfield  {author} {\bibinfo {author} {\bibfnamefont
  {M.}~\bibnamefont {White}}, \bibinfo {author} {\bibfnamefont
  {M.}~\bibnamefont {Pasienski}}, \bibinfo {author} {\bibfnamefont
  {D.}~\bibnamefont {McKay}}, \bibinfo {author} {\bibfnamefont {S.~Q.}\
  \bibnamefont {Zhou}}, \bibinfo {author} {\bibfnamefont {D.}~\bibnamefont
  {Ceperley}}, \ and\ \bibinfo {author} {\bibfnamefont {B.}~\bibnamefont
  {DeMarco}},\ }\bibfield  {title} {\enquote {\bibinfo {title} {Strongly
  interacting bosons in a disordered optical lattice},}\ }\href {\doibase
  10.1103/PhysRevLett.102.055301} {\bibfield  {journal} {\bibinfo  {journal}
  {Phys. Rev. Lett.}\ }\textbf {\bibinfo {volume} {102}},\ \bibinfo {pages}
  {055301} (\bibinfo {year} {2009})}\BibitemShut {NoStop}%
\bibitem [{\citenamefont {Paiva}\ \emph {et~al.}(2015)\citenamefont {Paiva},
  \citenamefont {Khatami}, \citenamefont {Yang}, \citenamefont {Rousseau},
  \citenamefont {Jarrell}, \citenamefont {Moreno}, \citenamefont {Hulet},\ and\
  \citenamefont {Scalettar}}]{Paiva2015}%
  \BibitemOpen
  \bibfield  {author} {\bibinfo {author} {\bibfnamefont {Thereza}\ \bibnamefont
  {Paiva}}, \bibinfo {author} {\bibfnamefont {Ehsan}\ \bibnamefont {Khatami}},
  \bibinfo {author} {\bibfnamefont {Shuxiang}\ \bibnamefont {Yang}}, \bibinfo
  {author} {\bibfnamefont {Val\'ery}\ \bibnamefont {Rousseau}}, \bibinfo
  {author} {\bibfnamefont {Mark}\ \bibnamefont {Jarrell}}, \bibinfo {author}
  {\bibfnamefont {Juana}\ \bibnamefont {Moreno}}, \bibinfo {author}
  {\bibfnamefont {Randall~G.}\ \bibnamefont {Hulet}}, \ and\ \bibinfo {author}
  {\bibfnamefont {Richard~T.}\ \bibnamefont {Scalettar}},\ }\bibfield  {title}
  {\enquote {\bibinfo {title} {Cooling atomic gases with disorder},}\ }\href
  {\doibase 10.1103/PhysRevLett.115.240402} {\bibfield  {journal} {\bibinfo
  {journal} {Phys. Rev. Lett.}\ }\textbf {\bibinfo {volume} {115}},\ \bibinfo
  {pages} {240402} (\bibinfo {year} {2015})}\BibitemShut {NoStop}%
\bibitem [{\citenamefont {Petrov}\ \emph {et~al.}(2000)\citenamefont {Petrov},
  \citenamefont {Holzmann},\ and\ \citenamefont {Shlyapnikov}}]{Petrov2000}%
  \BibitemOpen
  \bibfield  {author} {\bibinfo {author} {\bibfnamefont {D.~S.}\ \bibnamefont
  {Petrov}}, \bibinfo {author} {\bibfnamefont {M.}~\bibnamefont {Holzmann}}, \
  and\ \bibinfo {author} {\bibfnamefont {G.~V.}\ \bibnamefont {Shlyapnikov}},\
  }\bibfield  {title} {\enquote {\bibinfo {title} {Bose-einstein condensation
  in quasi-2d trapped gases},}\ }\href {\doibase 10.1103/PhysRevLett.84.2551}
  {\bibfield  {journal} {\bibinfo  {journal} {Phys. Rev. Lett.}\ }\textbf
  {\bibinfo {volume} {84}},\ \bibinfo {pages} {2551--2555} (\bibinfo {year}
  {2000})}\BibitemShut {NoStop}%
\bibitem [{\citenamefont {Huang}\ and\ \citenamefont {Meng}(1992)}]{Huang1992}%
  \BibitemOpen
  \bibfield  {author} {\bibinfo {author} {\bibfnamefont {Kerson}\ \bibnamefont
  {Huang}}\ and\ \bibinfo {author} {\bibfnamefont {Hsin-Fei}\ \bibnamefont
  {Meng}},\ }\bibfield  {title} {\enquote {\bibinfo {title} {Hard-sphere bose
  gas in random external potentials},}\ }\href {\doibase
  10.1103/PhysRevLett.69.644} {\bibfield  {journal} {\bibinfo  {journal} {Phys.
  Rev. Lett.}\ }\textbf {\bibinfo {volume} {69}},\ \bibinfo {pages} {644--647}
  (\bibinfo {year} {1992})}\BibitemShut {NoStop}%
\bibitem [{\citenamefont {Astrakharchik}\ \emph {et~al.}(2002)\citenamefont
  {Astrakharchik}, \citenamefont {Boronat}, \citenamefont {Casulleras},\ and\
  \citenamefont {Giorgini}}]{Astrakharchik2002}%
  \BibitemOpen
  \bibfield  {author} {\bibinfo {author} {\bibfnamefont {G.~E.}\ \bibnamefont
  {Astrakharchik}}, \bibinfo {author} {\bibfnamefont {J.}~\bibnamefont
  {Boronat}}, \bibinfo {author} {\bibfnamefont {J.}~\bibnamefont {Casulleras}},
  \ and\ \bibinfo {author} {\bibfnamefont {S.}~\bibnamefont {Giorgini}},\
  }\bibfield  {title} {\enquote {\bibinfo {title} {Superfluidity versus
  bose-einstein condensation in a bose gas with disorder},}\ }\href {\doibase
  10.1103/PhysRevA.66.023603} {\bibfield  {journal} {\bibinfo  {journal} {Phys.
  Rev. A}\ }\textbf {\bibinfo {volume} {66}},\ \bibinfo {pages} {023603}
  (\bibinfo {year} {2002})}\BibitemShut {NoStop}%
\bibitem [{\citenamefont {Yao}\ \emph {et~al.}(2020)\citenamefont {Yao},
  \citenamefont {Giamarchi},\ and\ \citenamefont {Sanchez-Palencia}}]{Yao2020}%
  \BibitemOpen
  \bibfield  {author} {\bibinfo {author} {\bibfnamefont {Hepeng}\ \bibnamefont
  {Yao}}, \bibinfo {author} {\bibfnamefont {Thierry}\ \bibnamefont
  {Giamarchi}}, \ and\ \bibinfo {author} {\bibfnamefont {Laurent}\ \bibnamefont
  {Sanchez-Palencia}},\ }\bibfield  {title} {\enquote {\bibinfo {title}
  {Lieb-liniger bosons in a shallow quasiperiodic potential: Bose glass phase
  and fractal mott lobes},}\ }\href {\doibase 10.1103/PhysRevLett.125.060401}
  {\bibfield  {journal} {\bibinfo  {journal} {Phys. Rev. Lett.}\ }\textbf
  {\bibinfo {volume} {125}},\ \bibinfo {pages} {060401} (\bibinfo {year}
  {2020})}\BibitemShut {NoStop}%
\bibitem [{\citenamefont {Peng}\ \emph {et~al.}(2010)\citenamefont {Peng},
  \citenamefont {Bohloul}, \citenamefont {Liu}, \citenamefont {Hu},\ and\
  \citenamefont {Drummond}}]{Peng2010}%
  \BibitemOpen
  \bibfield  {author} {\bibinfo {author} {\bibfnamefont {Shi-Guo}\ \bibnamefont
  {Peng}}, \bibinfo {author} {\bibfnamefont {Seyyed~S.}\ \bibnamefont
  {Bohloul}}, \bibinfo {author} {\bibfnamefont {Xia-Ji}\ \bibnamefont {Liu}},
  \bibinfo {author} {\bibfnamefont {Hui}\ \bibnamefont {Hu}}, \ and\ \bibinfo
  {author} {\bibfnamefont {Peter~D.}\ \bibnamefont {Drummond}},\ }\bibfield
  {title} {\enquote {\bibinfo {title} {Confinement-induced resonance in
  quasi-one-dimensional systems under transversely anisotropic confinement},}\
  }\href {\doibase 10.1103/PhysRevA.82.063633} {\bibfield  {journal} {\bibinfo
  {journal} {Phys. Rev. A}\ }\textbf {\bibinfo {volume} {82}},\ \bibinfo
  {pages} {063633} (\bibinfo {year} {2010})}\BibitemShut {NoStop}%
\bibitem [{\citenamefont {Zhang}\ and\ \citenamefont
  {Zhang}(2011)}]{Zhang2011}%
  \BibitemOpen
  \bibfield  {author} {\bibinfo {author} {\bibfnamefont {Wei}\ \bibnamefont
  {Zhang}}\ and\ \bibinfo {author} {\bibfnamefont {Peng}\ \bibnamefont
  {Zhang}},\ }\bibfield  {title} {\enquote {\bibinfo {title}
  {Confinement-induced resonances in quasi-one-dimensional traps with
  transverse anisotropy},}\ }\href {\doibase 10.1103/PhysRevA.83.053615}
  {\bibfield  {journal} {\bibinfo  {journal} {Phys. Rev. A}\ }\textbf {\bibinfo
  {volume} {83}},\ \bibinfo {pages} {053615} (\bibinfo {year}
  {2011})}\BibitemShut {NoStop}%
\bibitem [{\citenamefont {Bergeman}\ \emph {et~al.}(2003)\citenamefont
  {Bergeman}, \citenamefont {Moore},\ and\ \citenamefont
  {Olshanii}}]{Bergeman2003}%
  \BibitemOpen
  \bibfield  {author} {\bibinfo {author} {\bibfnamefont {T.}~\bibnamefont
  {Bergeman}}, \bibinfo {author} {\bibfnamefont {M.~G.}\ \bibnamefont {Moore}},
  \ and\ \bibinfo {author} {\bibfnamefont {M.}~\bibnamefont {Olshanii}},\
  }\bibfield  {title} {\enquote {\bibinfo {title} {Atom-atom scattering under
  cylindrical harmonic confinement: Numerical and analytic studies of the
  confinement induced resonance},}\ }\href {\doibase
  10.1103/PhysRevLett.91.163201} {\bibfield  {journal} {\bibinfo  {journal}
  {Phys. Rev. Lett.}\ }\textbf {\bibinfo {volume} {91}},\ \bibinfo {pages}
  {163201} (\bibinfo {year} {2003})}\BibitemShut {NoStop}%
\bibitem [{\citenamefont {Souza}\ \emph {et~al.}(2021)\citenamefont {Souza},
  \citenamefont {Pelster},\ and\ \citenamefont {dos Santos}}]{Souza2021}%
  \BibitemOpen
  \bibfield  {author} {\bibinfo {author} {\bibfnamefont {R~S}\ \bibnamefont
  {Souza}}, \bibinfo {author} {\bibfnamefont {Axel}\ \bibnamefont {Pelster}}, \
  and\ \bibinfo {author} {\bibfnamefont {F~E~A}\ \bibnamefont {dos Santos}},\
  }\bibfield  {title} {\enquote {\bibinfo {title} {Green's function approach to
  the bose{\textendash}hubbard model with disorder},}\ }\href {\doibase
  10.1088/1367-2630/ac15b3} {\bibfield  {journal} {\bibinfo  {journal} {New
  Journal of Physics}\ }\textbf {\bibinfo {volume} {23}},\ \bibinfo {pages}
  {083007} (\bibinfo {year} {2021})}\BibitemShut {NoStop}%
\bibitem [{\citenamefont {Lopatin}\ and\ \citenamefont
  {Vinokur}(2002)}]{Lopatin2002}%
  \BibitemOpen
  \bibfield  {author} {\bibinfo {author} {\bibfnamefont {A.~V.}\ \bibnamefont
  {Lopatin}}\ and\ \bibinfo {author} {\bibfnamefont {V.~M.}\ \bibnamefont
  {Vinokur}},\ }\bibfield  {title} {\enquote {\bibinfo {title} {Thermodynamics
  of the superfluid dilute bose gas with disorder},}\ }\href {\doibase
  10.1103/PhysRevLett.88.235503} {\bibfield  {journal} {\bibinfo  {journal}
  {Phys. Rev. Lett.}\ }\textbf {\bibinfo {volume} {88}},\ \bibinfo {pages}
  {235503} (\bibinfo {year} {2002})}\BibitemShut {NoStop}%
\bibitem [{\citenamefont {Xu}\ \emph {et~al.}(2006)\citenamefont {Xu},
  \citenamefont {Liu}, \citenamefont {Miller}, \citenamefont {Chin},
  \citenamefont {Setiawan},\ and\ \citenamefont {Ketterle}}]{Xu2006}%
  \BibitemOpen
  \bibfield  {author} {\bibinfo {author} {\bibfnamefont {K.}~\bibnamefont
  {Xu}}, \bibinfo {author} {\bibfnamefont {Y.}~\bibnamefont {Liu}}, \bibinfo
  {author} {\bibfnamefont {D.~E.}\ \bibnamefont {Miller}}, \bibinfo {author}
  {\bibfnamefont {J.~K.}\ \bibnamefont {Chin}}, \bibinfo {author}
  {\bibfnamefont {W.}~\bibnamefont {Setiawan}}, \ and\ \bibinfo {author}
  {\bibfnamefont {W.}~\bibnamefont {Ketterle}},\ }\bibfield  {title} {\enquote
  {\bibinfo {title} {Observation of strong quantum depletion in a gaseous
  bose-einstein condensate},}\ }\href {\doibase 10.1103/PhysRevLett.96.180405}
  {\bibfield  {journal} {\bibinfo  {journal} {Phys. Rev. Lett.}\ }\textbf
  {\bibinfo {volume} {96}},\ \bibinfo {pages} {180405} (\bibinfo {year}
  {2006})}\BibitemShut {NoStop}%
\bibitem [{\citenamefont {Slater}(1952)}]{Mathieu}%
  \BibitemOpen
  \bibfield  {author} {\bibinfo {author} {\bibfnamefont {J.~C.}\ \bibnamefont
  {Slater}},\ }\bibfield  {title} {\enquote {\bibinfo {title} {A soluble
  problem in energy bands},}\ }\href {\doibase 10.1103/PhysRev.87.807}
  {\bibfield  {journal} {\bibinfo  {journal} {Phys. Rev.}\ }\textbf {\bibinfo
  {volume} {87}},\ \bibinfo {pages} {807--835} (\bibinfo {year}
  {1952})}\BibitemShut {NoStop}%
\bibitem [{\citenamefont {Likharev}\ and\ \citenamefont {Zorin}(1985)}]{TL}%
  \BibitemOpen
  \bibfield  {author} {\bibinfo {author} {\bibfnamefont {K.~K.}\ \bibnamefont
  {Likharev}}\ and\ \bibinfo {author} {\bibfnamefont {A~.B.}\ \bibnamefont
  {Zorin}},\ }\bibfield  {title} {\enquote {\bibinfo {title} {Theory of the
  bloch-wave oscillations in small josephson junctions},}\ }\href {\doibase
  10.1007/BF00683782} {\bibfield  {journal} {\bibinfo  {journal} {J. Low. Temp.
  Phys.}\ }\textbf {\bibinfo {volume} {59}},\ \bibinfo {pages} {347} (\bibinfo
  {year} {1985})}\BibitemShut {NoStop}%
\bibitem [{\citenamefont {Du}\ \emph {et~al.}(2010)\citenamefont {Du},
  \citenamefont {Wan}, \citenamefont {Yesilada}, \citenamefont {Ryu},
  \citenamefont {Heinzen}, \citenamefont {Liang},\ and\ \citenamefont
  {Wu}}]{Du2010}%
  \BibitemOpen
  \bibfield  {author} {\bibinfo {author} {\bibfnamefont {X}~\bibnamefont {Du}},
  \bibinfo {author} {\bibfnamefont {Shoupu}\ \bibnamefont {Wan}}, \bibinfo
  {author} {\bibfnamefont {Emek}\ \bibnamefont {Yesilada}}, \bibinfo {author}
  {\bibfnamefont {C}~\bibnamefont {Ryu}}, \bibinfo {author} {\bibfnamefont
  {D~J}\ \bibnamefont {Heinzen}}, \bibinfo {author} {\bibfnamefont {Zhaoxin}\
  \bibnamefont {Liang}}, \ and\ \bibinfo {author} {\bibfnamefont {Biao}\
  \bibnamefont {Wu}},\ }\bibfield  {title} {\enquote {\bibinfo {title} {Bragg
  spectroscopy of a superfluid bose{\textendash}hubbard gas},}\ }\href
  {\doibase 10.1088/1367-2630/12/8/083025} {\bibfield  {journal} {\bibinfo
  {journal} {New Journal of Physics}\ }\textbf {\bibinfo {volume} {12}},\
  \bibinfo {pages} {083025} (\bibinfo {year} {2010})}\BibitemShut {NoStop}%
\end{thebibliography}%
\end{document}